\documentclass[journal]{IEEEtran}
\usepackage{graphicx}
\usepackage{cite}
\usepackage{picinpar}
\usepackage{amsmath}
\usepackage{url}
\usepackage[latin1]{inputenc}
\usepackage{colortbl}
\usepackage{soul}
\usepackage{multirow}
\usepackage{pifont}
\usepackage{color}
\usepackage{alltt}
\usepackage[hidelinks]{hyperref}
\usepackage{enumerate}
\usepackage{siunitx}
\usepackage{breakurl}
\usepackage{epstopdf}
\usepackage{pbox}

\usepackage{amssymb}
\usepackage{algorithm}
\usepackage{algorithmic}
\newcommand{\tabincell}[2]{\begin{tabular}{@{}#1@{}}#2\end{tabular}}
\allowdisplaybreaks[1]
\usepackage{amsthm}
\theoremstyle{remark}
\newtheorem{rem}{Remark}

\usepackage{ifpdf}
\usepackage{algorithmic}
\usepackage{array}
\ifCLASSOPTIONcompsoc
  \usepackage[caption=false,font=normalsize,labelfont=sf,textfont=sf]{subfig}
\else
 \usepackage[caption=false,font=footnotesize]{subfig}
\fi
\usepackage{fixltx2e}
\usepackage{stfloats}
\begin{document}
\title{A real-time distributed post-disaster restoration planning strategy for distribution networks}


\author{
	\vskip 1em
	
	Jianfeng~Fu,
	Alfredo~N\'u\~nez, \emph{Senior Member,~IEEE,} and~Bart~De~Schutter, \emph{Fellow,~IEEE}

	\thanks{
	
		Manuscript received Month xx, 2xxx; revised Month xx, xxxx; accepted Month x, xxxx.
		
		This work was supported by CSC (China Scholarship Council) with funding number: 201806280023.
		
		Jianfeng Fu and Bart De Schutter are with the Delft Center for Systems and Control, Delft University of Technology, Delft, The Netherlands, (e-mail: J.Fu-1@tudelft.nl, B.DeSchutter@tudelft.nl). 
		
		Alfredo~N\'u\~nez is with the Department of Engineering Structures, Delft University of Technology, (e-mail:A.A.NunezVicencio@tudelft.nl).
	}
}

\markboth{IEEE Transactions on Smart Grid,~Vol.~X, No.~X, April~2021}%
{Fu \MakeLowercase{\textit{et al.}}: A real-time distributed post-disaster restoration planning strategy for distribution networks}

\maketitle
	
\begin{abstract}
After disasters, distribution networks have to be restored by repair, reconfiguration, and power dispatch. During the restoration process, changes can occur in real time that deviate from the situations considered in pre-designed planning strategies. That may result in the pre-designed plan to become far from optimal or even unimplementable. This paper proposes a centralized-distributed bi-level optimization method to solve the real-time restoration planning problem. The first level determines integer variables related to routing of the crews and the status of the switches using a genetic algorithm (GA), while the second level determines the dispatch of active/reactive power by using distributed model predictive control (DMPC). A novel Aitken-DMPC solver is proposed to accelerate convergence and to make the method suitable for real-time decision making. A case study based on the IEEE 123-bus system is considered, and the acceleration performance of the proposed Aitken-DMPC solver is evaluated and compared with the standard DMPC method.
\end{abstract}


\section{Introduction}
Due to natural disasters, e.g., hurricanes, floods, and earthquakes, damages may emerge in the distributed network and result in outages. Thus, when recovering from disasters, an efficient restoration planning strategy is vital to accelerate the restoration of devices and to reduce load loss costs \cite{7915694,7893706,7922501}.\\
\indent In the literature, methods for optimal routing of repair crews, reconfiguration of the network, and dispatch of distributed generators (DGs) have been proposed to reduce the load loss costs in the restoration process \cite{7812566,8625548,8640043,8587140}. In \cite{7812566}, a two-stage method for the outage management of distribution networks is proposed. Based on clustering, a restoration planning method determines routes of the repair crews, reconfiguration, and DG dispatch. A coordination post-disaster repair planning method is proposed in \cite{8625548}, considering both the power system and also the natural gas system. In \cite{8640043}, the restoration process is modeled as a multi-crew scheduling problem with soft precedence constraints. A synthetic restoration planning model that integrates the reconfiguration with the crew routing process is proposed in \cite{8587140}.\\
\indent The restoration strategies mentioned above are pre-designed strategies that determine the global restoration plans for the whole restoration process after a disaster. However, the dynamic situations during the actual restoration process might differ from the situations assumed in the pre-designed restoration plans. Thus, the pre-designed plans may become far from optimal and even unimplementable in practice. For example, when facing newly emerged damages or dynamic traffic congestion, the pre-designed restoration planning methods cannot include crucial information that requires updates in real time. In \cite{8409997}, some of the uncertainties faced in restoration processes are included explicitly, such as demand and repair time. However, not all the uncertainties can be predicted properly.\\
\indent Thus, the current paper proposes a real-time restoration planning strategy to dynamically adapt to real-time situations that require an updated restoration plan to handle them properly. Besides, to avoid short-sighted planning results, the proposed restoration plan will be determined over a prediction window and implemented using a receding horizon strategy.\\
\indent The effectiveness of real-time planning strategies is highly related to computation speed. In dense and large-scale networks, real-time solutions can be challenging to obtain. Thus, a distributed model predictive control (DMPC) strategy that divides the large-scale planning problem into smaller-scale problems that communicate with each other is adopted. The smaller-scale planning problems can be solved in parallel, so that the computational burden can be further reduced \cite{2009Architectures,2013Distributed}. In previous works, for other applications in power systems, DMPC strategies have been applied to control, e.g., heat and electricity networks \cite{2014Distributed} and microgrids \cite{2019Hierarchical}. A DMPC strategy coordinates the solutions of the subsystems (smaller problems) to obtain a network-wide optimal solution. Different methods have been proposed for this coordination, e.g., Lagrangian multiplier methods \cite{2017Distributed}, Nash equilibrium methods \cite{2019Hierarchical}, Jacobi iteration methods \cite{5530634}, alternating direction method of multipliers \cite{8186925}. However, these methods cannot be implemented directly in planning problems, because they are designed to work with continuous variables \cite{2017Distributed,2019Hierarchical,5530634,8186925}, or because some of them (e.g., the alternating direction method of multipliers) can be implemented only for equality constraints \cite{8186925}. The planning problem in this paper includes both continuous and integer variables. In the literature, methodologies based on fixing integer variables and then using the DMPC strategy for solving the remaining continuous variables have been proposed in \cite{2014Distributed,mendes2017practical,Negenbornphd}. In the current paper, a similar strategy by fixing integer variables is followed but with a different fixing strategy (GA based) and a different solution strategy (Aitken-DMPC based) for solving the remaining problem after fixing the integer variables.\\
\indent More specifically, this paper proposes a centralized-distributed bi-level optimization architecture, whose higher level is centralized while the lower level is distributed. At the higher level, the chromosomes (integer solutions) generated by the genetic algorithm (GA) can be naturally processed in parallel, which accelerates the solution process. Then, the lower level utilizes an augmented-Lagrangian approach to coordinate the restoration plans of the subsystems so as to reach a network-wide optimal solution. At the lower level, a novel Aitken-DMPC solver is proposed to accelerate the convergence speed based on the Aitken algorithm \cite{lansky1992acceleration}. Furthermore, a warm start method is proposed to accelerate the solving speed.\\
\indent The contributions of this paper w.r.t. literature are:
\begin{enumerate}[1)]
\item A real-time restoration planning strategy for active distribution networks is proposed. Consequently, unpredictable situations can be addressed in real time so that the dynamic restoration plans can optimally reduce the load loss costs. 
\item A novel centralized-distributed bi-level optimization architecture is proposed to obtain solutions in real time. A GA-based approach is proposed to determine the integer variables at the higher level, and a DMPC solver is adopted to obtain the continuous variables at the lower level.
\item A novel Aitken-DMPC solver is proposed to accelerate the convergence to the optimal global solution.
\end{enumerate}
\indent The paper is organized as follows. Section II formulates the network-wide real-time restoration planning problem. Then, the proposed centralized-distributed bi-level optimization architecture is developed in Section III. In Section IV, the GA process and the Aitken-DMPC strategy are formalized. In Section V, a case study based on the IEEE 123-bus distribution network is presented. Finally, in Section VI, conclusions and further research are discussed.
\section{Real-time restoration problem}
\subsection{Real-time restoration based on receding horizon}
The restoration plans determined in real time may differ from the pre-designed plans as they will be adapting based on unpredictable events occurring while the restoration is being conducted. In that way, the effects of these unpredictable events can be reduced as much as possible.\\
\indent In this paper, a receding horizon procedure is adopted. Under this procedure, the power system operator determines the restoration plan over a prediction window according to the real-time statuses that include the current locations of the repair crews, the travel time for crews to reach the damages, and the locations of new damages. Then, the restoration plan for the current time step covering a fixed time period (referred to as prediction window from now on) is reassessed and executed. Repair crews travel from their current locations to the damages, the switches are operated to reconfigure the network, and the powers generated by DGs are dispatched so that the load loss costs in the network can be minimized. At the next time step, the procedure is repeated, the real-time statuses are updated, the prediction window is shifted with one time step, and a new optimization problem is solved to determine an updated restoration plan.
\subsection{Network-wide restoration problem formulation}
A general distribution network with switches, DGs, loads, and transmission lines is considered. When the network is affected by disasters, the restoration strategy determines repair crew routing, reconfiguration, and dispatching of DGs. The objective function of the restoration planning problem is defined as:
\begin{equation}
\min J=\sum\limits_{t\in\mathcal{T}}\sum\limits_{n\in\mathcal{N}} c_n(t)\cdot (P^{\rm L}_{{\rm max},n}(t)-P^{\rm L}_n(t)) \label{3a}\tag{3a}
\end{equation}
where $\mathcal{T}$ is the set of time steps in the prediction window, $\mathcal{N}$ is the set of buses in the network, $c_n(t)$ is the load loss cost that reveals the importance of the load on bus $n$ at time step $t$, $P^{\rm L}_{{\rm max},n}(t)$ is the predicted load demand on bus $n$ at time step $t$, and $P^{\rm L}_n(t)$ is the actual load on bus $n$ at time step $t$. Besides, the following constraints are considered:
\begin{gather}
0\leq P^{\rm DG}_n(t)\leq P^{\rm DG}_{{\rm max},n}(t),\forall n\in\mathcal{N},\,\forall t\in\mathcal{T}\label{3b}\tag{3b}\\
0\leq Q^{\rm DG}_n(t)\leq Q^{\rm DG}_{{\rm max},n}(t),\forall n\in\mathcal{N},\,\forall t\in\mathcal{T}\label{3c}\tag{3c}\\
0\leq P^{\rm L}_n(t)\leq P^{\rm L}_{{\rm max},n}(t),\forall n\in\mathcal{N},\,\forall t\in\mathcal{T}\label{3d}\tag{3d}\\
0\leq Q^{\rm L}_n(t)\leq Q^{\rm L}_{{\rm max},n}(t),\forall n\in\mathcal{N},\,\forall t\in\mathcal{T}\label{3e}\tag{3e}\\
-\delta_{n,m}(t) P^{\rm Tr}_{{\rm max},n,m}\leq P^{\rm Tr}_{n,m}(t)\leq \delta_{n,m}(t)P^{\rm Tr}_{{\rm max},n,m},\notag\\
\forall m\in\mathcal{M}_{n},\,\forall n\in\mathcal{N},\forall t\in\mathcal{T}\label{3f}\tag{3f}\\
-\delta_{n,m}(t) Q^{\rm Tr}_{{\rm max},n,m}\leq Q^{\rm Tr}_{n,m}(t)\leq \delta_{n,m}(t)Q^{\rm Tr}_{{\rm max},n,m},\notag\\
\forall m\in\mathcal{M}_{n},\,\forall n\in\mathcal{N},\forall t\in\mathcal{T}\label{3g}\tag{3g}\\
\delta_{n,m}(t)=1,\,\forall m\in\mathcal{M}_n,\,\forall (n,m)\notin\Omega_{\rm SW}\cup\Omega_{\rm DA},\,\forall t\in\mathcal{T}\label{3h}\tag{3h}\\
\sum\limits_{m\in\mathcal{M}^{\rm I}_{n}}P^{\rm Tr}_{m,n}(t)+P^{\rm DG}_n(t)=\sum\limits_{m\in\mathcal{M}^{\rm O}_{n}}P^{\rm Tr}_{n,m}(t)+P^{\rm L}_n(t),\notag\\
\forall n\in\mathcal{N},\,\forall t\in\mathcal{T}\label{3i}\tag{3i}\\
\sum\limits_{m\in\mathcal{M}^{\rm I}_{n}}Q^{\rm Tr}_{m,n}(t)+Q^{\rm DG}_n(t)=\sum\limits_{m\in\mathcal{M}^{\rm O}_{n}}Q^{\rm Tr}_{n,m}(t)+Q^{\rm L}_n(t),\notag\\
\forall n\in\mathcal{N},\,\forall t\in\mathcal{T}\label{3j}\tag{3j}\\
(\delta_{n,m}(t)-1)\cdot M\leq V_n(t)-V_m(t)+ \notag \\
(R_{n,m}P^{\rm Tr}_{n,m}(t)+X_{n,m} Q^{\rm Tr}_{n,m}(t))/V_{\rm ref}\leq (1-\delta_{n,m}(t))\cdot M,\notag\\ 
\forall n\in\mathcal{N},\forall m\in\mathcal{M}_{n} \label{3k}\tag{3k}\\
1-\varepsilon\leq V_n(t)/V_{\rm ref}\leq 1+\varepsilon,\,\forall n\in\mathcal{N},\,\forall t\in\mathcal{T} \label{3l}\tag{3l}
\end{gather} 
Constraints \eqref{3b} and \eqref{3c} indicate that the generation active power $P^{\rm DG}_n$ and reactive power $Q^{\rm DG}_n$ for bus $n$ at time step $t$ cannot exceed their maximum limits, $P^{\rm DG}_{{\rm max},n}$ and $Q^{\rm DG}_{{\rm max},n}$ respectively. Similarly, in constraints \eqref{3d} and \eqref{3e}, the generation active power $P^{\rm L}_n$ and reactive power $Q^{\rm L}_n$ of the loads on bus $n$ at time step $t$ are limited by their maximum levels, $P^{\rm L}_{{\rm max},n}$ and $Q^{\rm L}_{{\rm max},n}$ respectively. Constraints \eqref{3f} and \eqref{3g} represent the active and reactive power limitations of the transmission line between bus $n$ and its neighbor bus $m$, where $\delta_{n,m}(t)$ equals to 1 if bus $n$ and $m$ are connected (this means that neither the transmission line between bus $n$ and $m$ is damaged nor the switch on this transmission line is open) at time step $t$, and 0 otherwise, $\mathcal{M}_{n}$ is the set of neighbor buses of bus $n$, $P^{\rm Tr}_{{\rm max},n,m}$ and $Q^{\rm Tr}_{{\rm max},n,m}$ are the maximum active power and reactive power that the transmission line between bus $n$ and $m$ can transmit, $P^{\rm Tr}_{n,m}(t)$ and  $Q^{\rm Tr}_{n,m}(t)$ are respectively the transmission active power and reactive power between bus $n$ and $m$ at time step $t$. Constraint (3h) represents that when there is no switch nor damage on the line, neighbor buses $n$ and $m$ are always connected, where $\Omega_{\rm SW}$ and $\Omega_{\rm DA}$ are the sets of lines with switches nor damages. Constraints (3i) and (3j) represent the active power and reactive power balance of bus $n$ respectively, where $\mathcal{M}^{\rm I}_n$ and $\mathcal{M}^{\rm O}_n$ are sets of the neighbor buses that are injecting into bus $n$ or that are receiving an outflow from bus $n$ respectively. Constraint (3k) represents the voltage relationship between neighbor buses, where $R_{n,m}$, $X_{n,m}$ are the resistance and reactance between bus $n$ and its neighbor bus $m$, $V_n(t)$ is the voltage of bus $n$ at time step $t$, and $V_{\rm ref}$ is the reference voltage. Constraint (3l) limits the voltages to a relative deviation of at most $\varepsilon$ with respect to $V_{\rm ref}$ (a typical value of $\varepsilon$ is 0.05). The routing process of the crews satisfies the following constraints:
\begin{gather}
T_{p_x,c}=T_{p_x-1,c}+\tau_{p_x-1,p_x}(t)+\upsilon_{p_x-1,c},\,\forall c\in\mathcal{C}\,\,\,{\rm if}\,{\rm the}\notag\\
{\rm route}\,\,{\rm is}\,\, p_1\mapsto...\mapsto p_x-1\mapsto p_x\mapsto...\mapsto p_{r_c}\label{3m}\tag{3m}\\
\delta_{n_p,m_p}(t)=\begin{cases}
1 & \mbox{if $ T_{p,c}+\upsilon_{p,c}\leq t\cdot \Delta t$ and $\Delta_{n_p,m_p}(t)=1$}\\
0 & \mbox{otherwise}
\end{cases}\notag\\
\forall m_p\in\mathcal{M}_{n_p},\,(n_p,m_p)\in\Omega_{\rm SW}\cup\Omega_{\rm DA},\,\forall t\in\mathcal{T}\label{3n}\tag{3n}
\end{gather}
Constraint \eqref{3m} represents the routing of crew $c$ where $\mathcal{C}$ is the set of repair crews, $T_{p,c}$ is the time instant crew $c$ starts to repair damage $p$,  $\tau_{p,q,c}(t)$ is the duration taken for traveling between damage $p$ and $q$ for crew $c$ at time step $t$, $\upsilon_{p,c}$ is the repair time of damage $p$ for crew $c$, $n_p$ and $m_p$ are two terminal buses of the line on which damage $p$ occurs, and $\Delta_{m_p,n_p}(t)$ is equal to 1 if the switch on the transmission line between buses $n_p$ and $m_p$ is closed at time step $t$, and 0 otherwise. In \eqref{3m}, ``$\mapsto$'' is used to indicate the order under which the damages assigned to crew $c$ are fixed. Constraint \eqref{3n} indicates that if the damage $p$ is between bus $n_p$ and $m_p$ is fixed before the current time step $t$ and the switch between bus $n_p$ and $m_p$ is closed for time step $t$, then $\delta_{n_p,m_p}(t)$ should be equal to 1, where $\Delta t$ is the duration length of a time step.\\
\indent Different from the pre-designed restoration planning strategy, in general not all the damages have to be fixed in one prediction window. Besides, the on-going repair tasks that started before the current time step but that have not yet been completed should also be considered when planning for the prediction window corresponding to the current time step. These two problems will be tackled following the solution method in Section IV.A.
\section{Centralized-distributed bi-level optimization architecture}
\indent For a large-scale and complex distribution network, solving a restoration planning problem for the whole network is time prohibitive. Besides, real-time planning requires a fast solution speed. Therefore, a DMPC solver is adopted. The DMPC solver optimizes the system-wide restoration plan for the whole network. It first optimizes small-scale optimal restoration plans for each subsystem in parallel, and then interconnects the optimal plans for the subsystems to obtain a system-wide optimal plan.\\
\begin{figure}[t!]
\centering
\includegraphics[width=0.5\textwidth]{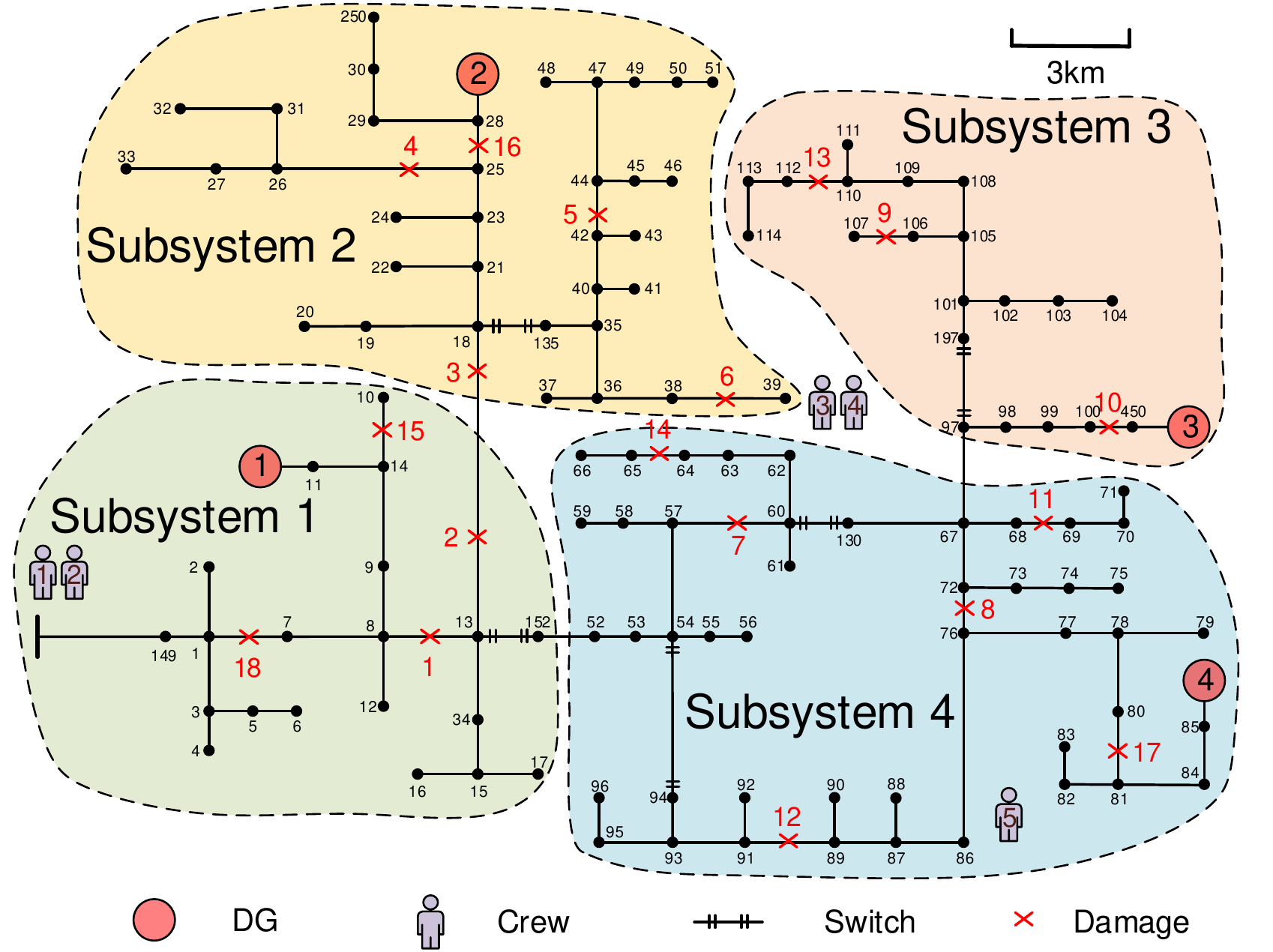}
\DeclareGraphicsExtensions.
\caption{An illustrative IEEE 123-bus network and its subsystems}
\label{fig_sim}
\end{figure}
\indent An illustrative example of the whole network and sub-systems is given in Fig. 1. The current locations of damages, the locations of the crews and depots, and the topology of the distribution network are assumed to be given. In Fig. 1, different subsystems are in areas with different colors and four subsystems are considered\footnote{The method to divide the overall system into subsystems is discussed in Remark 1.}. To enable distributed computation, this paper proposes a centralized-distributed bi-level optimization architecture that is shown in Fig. 2. In Fig. 2, the GA-based coordinator generates multiple chromosomes in one iteration and then distributes each chromosome (values of the fixed integer variables) to each system agent, and each system agent obtains the system-wide restoration plan according to the values of fixed integer variables by coordinating the subsystem-wide restoration plans of its subsystem agents.\\
\indent At the higher level of the architecture, in the GA-based coordinator, each chromosome corresponds to a combination of the values of all the integer variables in the whole network. More specifically, these include the open-closed statuses of the switches, and the route to visit the sequence of damages $p\mapsto...\mapsto p_{r_c}$ for crew $c\in\mathcal{C}$.\\
\indent In the lower level of the architecture, the values of integer variables are substituted into problem (3), and then, problem (3) becomes a continuous planning problem. So, each individual chromosome corresponds to one continuous planning problem and one system agent. Still, the continuous planning problem consists of a large number of constraints and variables. To reduce the computing time, a DMPC solver is adopted in the lower level of the architecture. For each system agent, DMPC is implemented by dividing the remaining continuous system planning problem into subsystem planning problems. Each subsystem-wide planning problem corresponds to one of the $n_s$ subsystem agents. Each subsystem agent optimizes its decoupled subsystem restoration planning problem, and then communicates with other subsystem agents to obtain the system-wide optimal plan (see Section IV.B). After each system agent obtains the system-wide optimal plan by coordination, the system-wide optimal solution will be transmitted back to the GA-based coordinator to determine the chromosomes for the next iteration by the collected optimal solutions from all the system agents.\\
\indent An advantage of the proposed centralized-distributed bi-level optimization architecture is that the solving processes of the system agents and subsystem agents can both be implemented in parallel. Consequently, the computation time can be largely reduced. For practical implementation considerations, it is not difficult for a power system operator to install a large number of CPUs in order to arrange hundreds of agents to complete the parallel computing tasks.
\begin{figure*}[t!]
\centering
\includegraphics[width=0.75\textwidth]{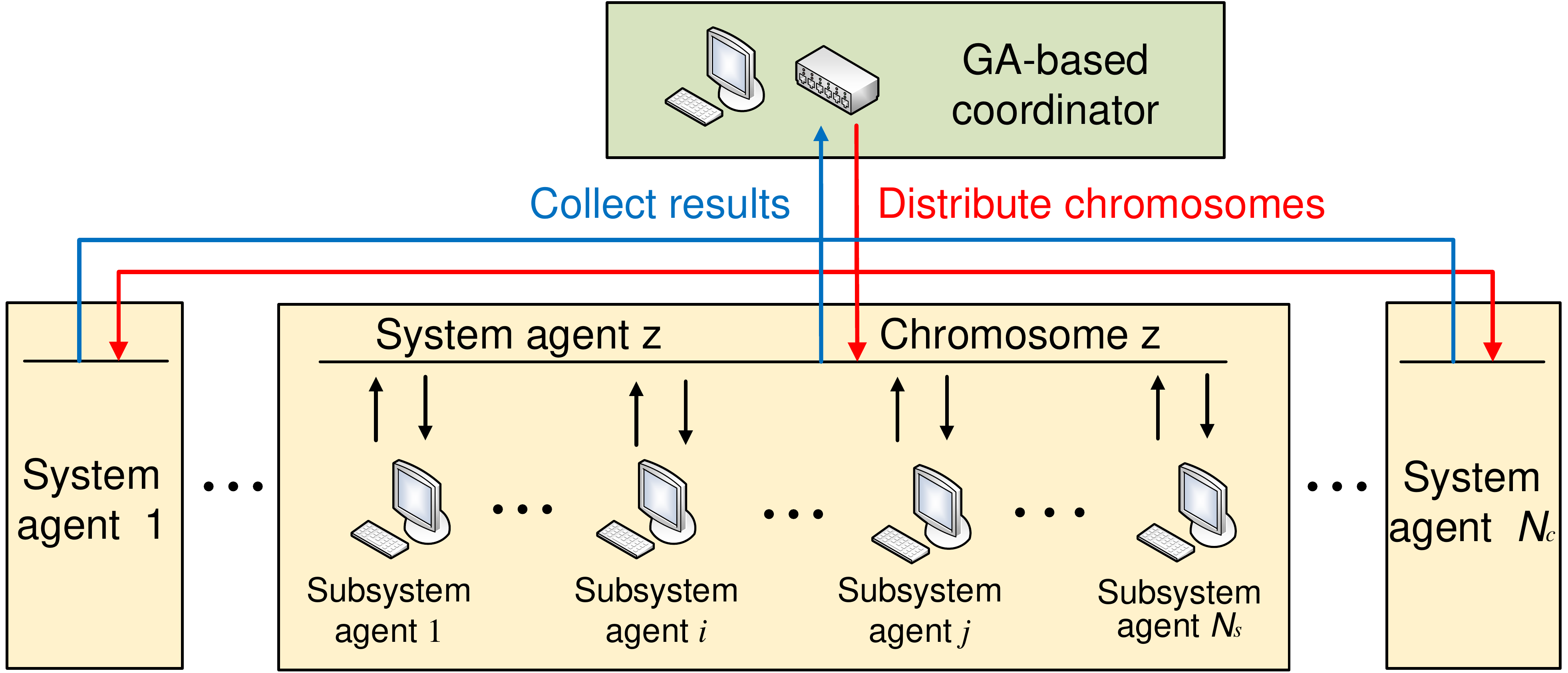}
\DeclareGraphicsExtensions.
\caption{Centralized-distributed bi-level optimization architecture}
\label{fig_sim}
\end{figure*}
\begin{rem}
When dividing the network into subsystems, there is a trade-off between two aspects. Firstly, if the scales of the planning problems of subsystems are different, the planning problems may require different computation times to be solved. The subsystems that solve their problems quicker will then have to wait for the others to finish. This is because the GA mutation and crossover process of the GA-based coordinator needs all the solutions. So the effective computation time for each generation of the GA is determined by the maximum computation time among all the subsystems. Secondly, the fewer interconnecting variables between the subsystems, the less time is needed for convergence \cite{Negenbornphd}. Thus, in the current paper, a partitioning method is selected that provides a similar number of variables for each subsystem while minimizing the number of interconnecting variables. In the literature, several partitioning methods to find adequate subsystems have been proposed. For instance in \cite{1525099}, a partitioning process including network coarsening, graph partitioning, and refining is proposed for partitioning the distributed computing network for power system transient stability simulation. The current paper adopts the partitioning method of \cite{1525099}. 
\end{rem}
\begin{rem}
\indent A warm start method is proposed to be applied at both levels: for the generation of initial individuals in the GA and for the initial values of the variables in the DMPC step. The restoration plans are obtained using a receding horizon procedure. As a consequence, the unimplemented restoration plans of the previous time step extended with a feasible action, ``\textit{Doing nothing}'', at the end of the plan, can be used as the warm start solution. In addition, an initial solution for the first time step can be obtained from the pre-designed restoration plan.

\end{rem}
\section{GA-based coordination and agents for DMPC}
\indent In this section, the optimization problems in the centralized-distributed bi-level optimization architecture will be formalized. The design of the GA-based coordination and the design of agents for DMPC implementation will be explained.
\subsection{Design of the GA-based coordination}
\indent Constraint \eqref{3l} includes the routing decisions of the restoration process and constraint \eqref{3n} includes the open-closed statuses of switches. To represent the routes and open-closed statuses of switches in chromosomes, the chromosomes are coded as a combination of integers and binary values, as shown in Fig. 3. A two-part chromosome representation is proposed when modeling the routes, where each box represents a gene. The first part of the chromosome is a sequence of damages. The second part represents the number of damages that the crews will fix. If there are $z$ crews, then $z-1$ genes are used, where each gene $l$, $l=1,2,...,z-1$, represents the number of damages of the corresponding crew, and the last crew will take the remaining damages. For instance, in Fig. 3, the routes for crews 1, 2, and 3 are $6\mapsto 4$, $2$, and $1\mapsto 3\mapsto 5$ respectively. In this paper, integer variables are adopted to represent the routes, but not binary variables as mostly modeled in the literature of the power system restoration routing strategies \cite{8409997}. That is because for binary coding with $n_{\rm b}$ bits, the number of feasible combinations might not always equal the total number of chromosomes $2^{n_{\rm b}}$ available. For instance, if there are 4 damages to be visited, then there are 24 possible combinations of sequences to visit them considering one single crew. To represent this number of routes, 5 bits are needed. Then if $n_{\rm b}$ is equal to 5, there will be 32 possible chromosomes. However, with the integer coding, each damage can be represented by one gene, and its position within the chromosomes indicates the order within the sequence. For modeling the open-closed statuses of the switches, the binary coding contains the open-closed statuses of the time steps in the prediction window. For instance in Fig. 3, the last six genes represent the open-closed statuses of two switches in a prediction window with three time steps. Switch 1 is open at time step 1 and 2 and closed at time step 3, while switch 2 is closed at time step 1 and 2 and open at time step 3. Then, the chromosomes in Fig. 3 are of the same length, with the following structure: (damages) \# (crews) \# (open-closed statuses for each time step), where \# represents concatenation. \\
\begin{figure}[!t]
\centering
\includegraphics[scale=0.6]{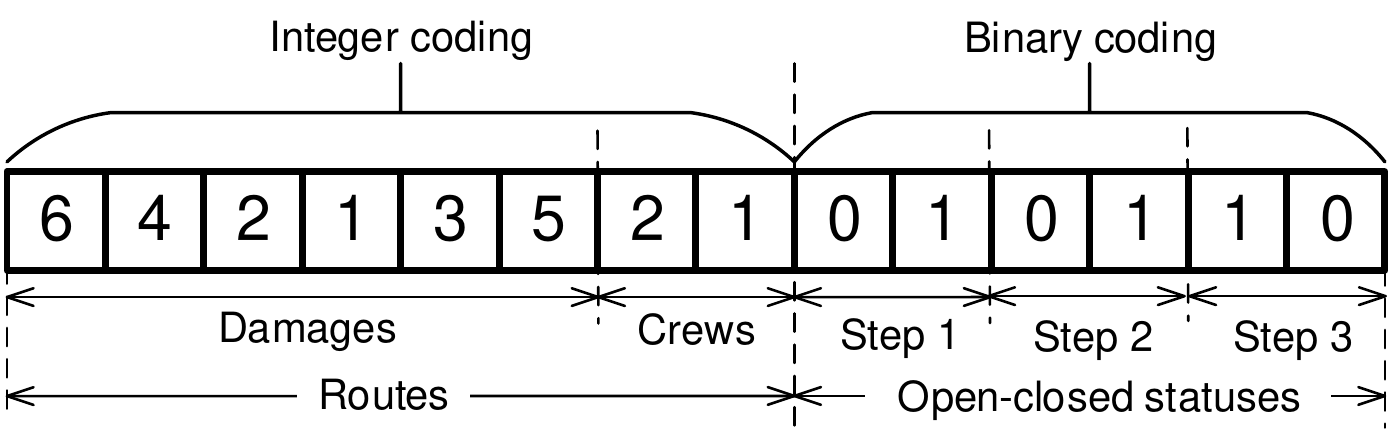}
\DeclareGraphicsExtensions.
\caption{An example of chromosome representation}
\label{fig_sim}
\end{figure}
\indent With the given representation of the chromosomes, at each time step, the chromosomes are generated by the GA-based coordination for the current prediction window. The method for generating the damages part of the chromosome in Fig. 3 for the next iteration is based on flipping, swapping, and sliding operators \cite{2000A}. It should be noticed that, different from the traveling salesman problem implemented in \cite{2000A}, the real-time restoration routing problem has a prediction window that will limit the number of damages that can be visited by the crews. So if a route extends beyond the end of the prediction window, only the part inside the prediction window will be considered to compute the fitness function. Furthermore, when determining the routes for the current time step, some crews may still be performing their repair tasks that started before the current time step, but these tasks might be completed within the current prediction window. When calculating the fitness function, for such crews only the part of the route starting from when the crews have finished their ongoing repair tasks is considered. The methods for generating the crews parts and the open-closed statuses parts of the chromosomes are the same as in traditional GA methods and they include crossover and mutation for integer coding and binary coding \cite{2006A}.\\
\indent Note that, after substituting each chromosome in (3), the remaining programming problem is always feasible for each chromosome, because at least there exists a feasible solution obtained by setting active powers and reactive powers to zero. Hence, there is no need for penalty terms for infeasibility on the fitness function of GA, and so the fitness function is the same as (3a).
\subsection{Design of the agents for DMPC implementation}
\indent Next, the optimization problem solved by the subsystem agents is formalized and it is explained how the optimal plans for the subsystems are coordinated to obtain an optimal system-wide plan. Fig. 4 shows a system topology after subsystem division. In Fig. 4, the interconnecting variables with its neighbor $j$ for subsystem $i$ are $\widetilde{P}_{{\rm in},ji}(t)$ (active power from subsystem $j$ to $i$), $\widetilde{Q}_{{\rm in},ji}(t)$ (reactive power from subsystem $j$ to $i$), and $\widetilde{V}_{{\rm in},ji}(t)$ (voltage of subsystem $i$ at the $i$ side terminal of the link between $i$ and $j$), while the interconnecting variables for subsystem $j$ are $\widetilde{P}_{{\rm out},ij}(t)$ (active power from subsystem $i$ to $j$), $\widetilde{Q}_{{\rm out},ij}(t)$ (reactive power from subsystem $i$ to $j$), and $\widetilde{V}_{{\rm out},ij}(t)$ (voltage of subsystem $j$ at the $j$ side terminal of the link between $i$ and $j$). The aim of the coordination approach of DMPC is to make the pairs of ``in'' and ``out'' values (e.g., $\widetilde{P}_{{\rm in},ji}(t)$ and $\widetilde{P}_{{\rm out},ij}(t)$) converge to an optimal value by iteration. In this paper, augmented-Lagrangian DMPC \cite{Negenbornphd} is adopted as the coordination approach. To accelerate the convergence speed of the DMPC iteration process, a novel idea that uses the Aitken algorithm, which is a second-order convergence algorithm \cite{lansky1992acceleration}, to update the Lagrangian multipliers is proposed.\\
\begin{figure}[!t]
\centering
\includegraphics[scale=0.5]{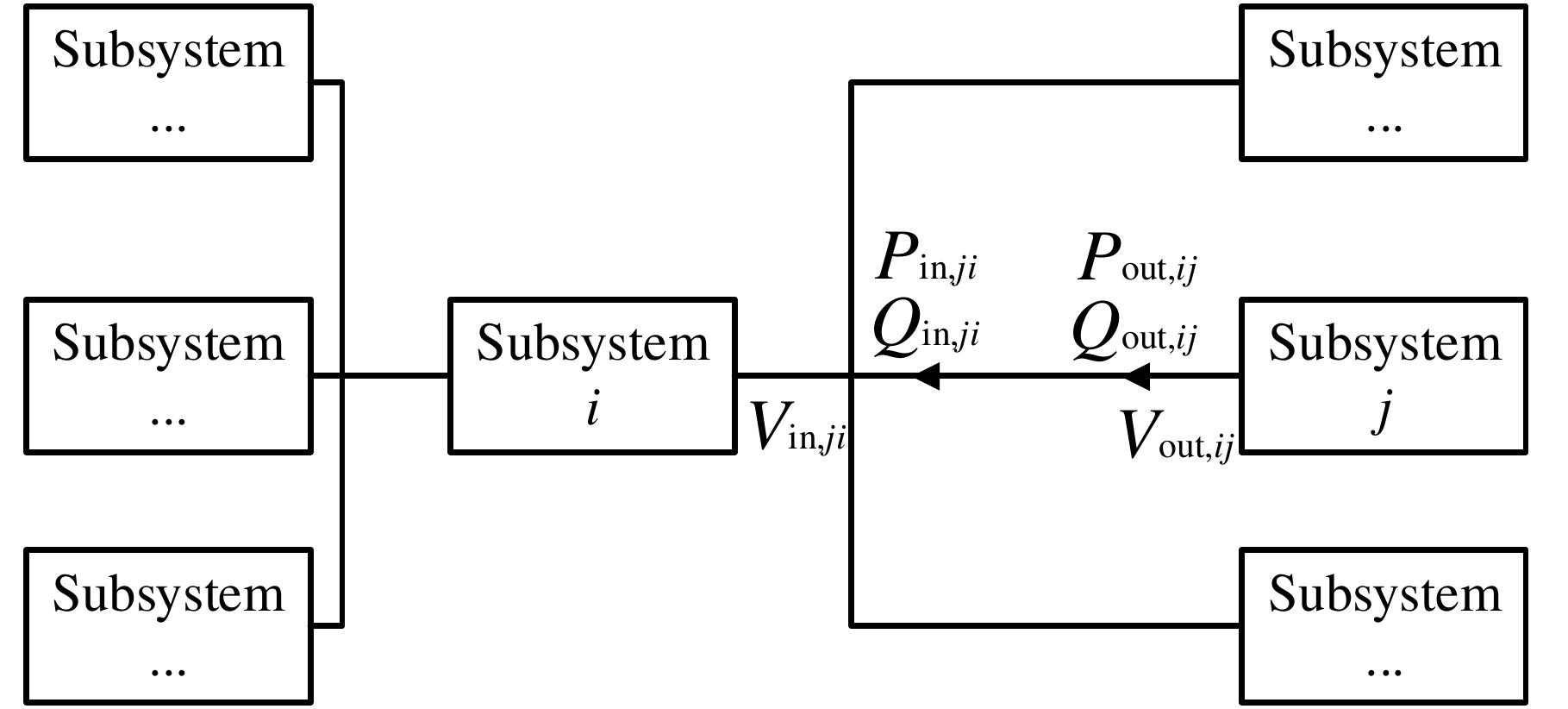}
\DeclareGraphicsExtensions.
\caption{An illustration for subsystems and interconnecting variables}
\label{fig_sim}
\end{figure}
\indent When the integer variables in (3) are fixed, the objective function of one subsystem can be represented as:
\begin{equation}
\begin{gathered}
J_{{\rm local},i}=\sum\limits_{t\in\mathcal{T}}\sum\limits_{n\in\mathcal{N}_i} c_n(t)\cdot (P^{\rm L}_{{\rm max},n}(t)-P^{\rm L}_n(t))\label{5}\tag{5}
\end{gathered}
\end{equation}
where $\mathcal{N}_i$ is the set of buses in subsystem $i$. To represent the communication between the subsystems, the augmented-Lagrangian term with quadratic terms is introduced for subsystem $i$ for iteration $k$ at time step $t$\cite{Negenbornphd}, such that:
\begin{equation}
\begin{gathered}
J_{{\rm inter},i}^{(k)}(t)=\sum\limits_{j\in\mathcal{I}_i}\Big(\widetilde{\lambda}^{(k)}_{{\rm in},ji}(t)\cdot\widetilde{P}_{{\rm in},ji}(t)+\widetilde{\mu}^{(k)}_{{\rm in},ji}(t)\cdot\\
\widetilde{Q}_{{\rm in},ji}(t)+\widetilde{\nu}^{(k)}_{{\rm in},ji}(t)\cdot \widetilde{V}_{{\rm in},ji}(t)+\frac{\gamma_c}{2}(\widetilde{P}_{{\rm out},{\rm prev},ij}(t)-\\
\widetilde{P}_{{\rm in},ji}(t))^2+\frac{\gamma_c}{2}\cdot(\widetilde{Q}_{{\rm out},{\rm prev},ij}(t)-\widetilde{Q}_{{\rm in},ji}(t))^2+\\
\frac{\gamma_c}{2}\cdot(\widetilde{V}_{{\rm out},{\rm prev},ij}(t)-\widetilde{V}_{{\rm in},ji}(t))^2+\frac{\gamma_b-\gamma_c}{2}(\widetilde{P}_{{\rm in},ji}(t)-\\
\widetilde{P}_{{\rm in},{\rm prev},ji}(t))^2+\frac{\gamma_b-\gamma_c}{2}(\widetilde{Q}_{{\rm in},ji}(t)-\widetilde{Q}_{{\rm in},{\rm prev},ji}(t))^2\Big)+\\
\frac{\gamma_b-\gamma_c}{2}(\widetilde{V}_{{\rm in},ji}(t)-\widetilde{V}_{{\rm in},{\rm prev},ji}(t))^2\Big)
\end{gathered}\label{6}\tag{6}
\end{equation}
where $\mathcal{I}_i$ is the set of neighbor subsystems of subsystem $i$, $\widetilde{\lambda}^{(k)}_{{\rm in},ji}(t)$, $\widetilde{\mu}^{(k)}_{{\rm in},ji}(t)$, and $\widetilde{\nu}^{(k)}_{{\rm in},ji}(t)$ are the Lagrangian multipliers for the interconnecting constraints $\widetilde{P}_{{\rm in},ji}(t)=\widetilde{P}_{{\rm out},ij}(t)$, $\widetilde{Q}_{{\rm in},ji}(t)=\widetilde{Q}_{{\rm out},ij}(t)$, and $\widetilde{V}_{{\rm in},ji}(t)=\widetilde{V}_{{\rm out},ij}(t)$ respectively, $\widetilde{P}_{{\rm out},{\rm prev},ij}(t)$ and $\widetilde{P}_{{\rm in},{\rm prev},ji}(t)$ are set equal to the values $\widetilde{P}_{{\rm out},ij}(t)$ and $\widetilde{P}_{{\rm in},ji}(t)$ obtained at iteration $k-1$ (i.e. the previous iteration), and similarly for $\widetilde{Q}_{{\rm out},{\rm prev},ji}(t)$ and $\widetilde{Q}_{{\rm in},{\rm prev},ji}(t)$, $\widetilde{V}_{{\rm out},{\rm prev},ji}(t)$, and $\widetilde{V}_{{\rm in},{\rm prev},ji}(t)$. The augmented-Lagrangian term is a quadratic term that makes the optimization problem strictly convex. Then the process for agents to solve their optimization problem at time step $t$ for one prediction window is designed as follows:
\subsubsection{Status measurement} At time step $t$, each agent $i$ measures the real-time statuses and their expected profiles over the prediction window, including DG profiles $P^{\rm DG}_{{\rm max},n}(\cdot)$ and $Q^{\rm DG}_{{\rm max},n}(\cdot)$, load profiles $P^{\rm L}_{{\rm max},n}(\cdot)$ and $Q^{\rm L}_{{\rm max},n}(\cdot)$, transmission limitations $P^{\rm Tr}_{\rm max}(\cdot)$ and $Q^{\rm Tr}_{\rm max}(\cdot)$, and parameters including traveling time $\tau$ and durations $\upsilon$ for fixing damages.
\subsubsection{Initialization} Set the iteration counter $k$ to 1. Initialize the Lagrangian multipliers according to Remark 2.
\subsubsection{Substituting and solving} Substitute the values of one chromosome into the formulated problem and for agent $i$ determine $\Omega^{(k)}(t)$=[$\widetilde{P}^{(k)}_{{\rm in},ji}(t)$,$\widetilde{Q}^{(k)}_{{\rm in},ji}(t)$,$\widetilde{V}^{(k)}_{{\rm in},ji}(t)$]$_{i\in\mathcal{I},j\in\mathcal{I}_i}$, $\Theta^{(k)}(t)$=[$P_n^{{\rm L},(k)}(t)$,$Q_n^{{\rm L},(k)}(t)$,$P_n^{{\rm DG},(k)}(t)$,$Q_n^{{\rm DG},(k)}(t)$,$P_{n,m}^{{\rm Tr},(k)}(t)$, $Q_{n,m}^{{\rm Tr},(k)}(t)$,$V_n^{(k)}(t)$]$_{n\in\mathcal{N}_i,m\in\mathcal{M}_{n,i}}$ where $\mathcal{M}_{n,i}$ is the set of neighbor buses of bus $n$ in subsystem $i$ with objective function:
\begin{equation}
\begin{gathered}
\min_{\Omega^{(k)}(t),\Theta^{(k)}(t)} J_{{\rm local},i}+\sum\limits_{t\in\mathcal{T}}J^{(k)}_{{\rm inter},i}(\widetilde{\lambda}^{(k)}_{{\rm in},ji}(t), \widetilde{\mu}^{(k)}_{{\rm in},ji}(t),\\
\widetilde{\nu}^{(k)}_{{\rm in},ji}(t),\widetilde{P}^{(k)}_{{\rm in},ji}(t), \widetilde{Q}^{(k)}_{{\rm in},ji}(t),\widetilde{V}^{(k)}_{{\rm in},ji}(t))\label{7}\tag{7}
\end{gathered}
\end{equation}
\indent The constraints of the problem solved by agent $i$ are the local constraints in (3) of subsystem $i$ over one prediction window. The agents solve their own problem in parallel and send the results to their system agents via communication.
\setcounter{subsubsection}{3}
\subsubsection{Update Lagrangian multipliers} Update Lagrange multipliers $\widetilde{\lambda}$, $\widetilde{\mu}$, and $\widetilde{\nu}$ in system agents by\footnote{In (8a)-(8c) and (9), only the multipliers $\widetilde{\lambda}$ for $\widetilde{P}$ are considered. The formulas for the multipliers $\widetilde{\mu}$ for $\widetilde{Q}$ and for the multipliers $\widetilde{\nu}$ for $\widetilde{V}$ are similar.}:
\begin{equation}
\begin{gathered}
\widetilde{\lambda}^{(k+1)}=\varphi(\widetilde{\lambda}^{(k)})=\widetilde{\lambda}^{(k)}+\gamma_c(\widetilde{P}_{{\rm in},ji}^{(k)}-\widetilde{P}_{{\rm out},ij}^{(k)})\label{8a}\tag{8a}
\end{gathered}
\end{equation}
Then iterate twice via $\varphi$ with the Aitken algorithm \cite{lansky1992acceleration}:
\begin{equation}
\begin{gathered}
\hat{\lambda}^{(k+1)}=\varphi(\widetilde{\lambda}^{(k)}),\,\overline{\lambda}^{(k+1)}=\varphi(\hat{\lambda}^{(k+1)})
\end{gathered}\label{8b}\tag{8b}
\end{equation}
Then calculate the Lagrangian multipliers for the next iteration:
\begin{equation}
\begin{gathered}
\widetilde{\lambda}^{(k+1)}=\overline{\lambda}^{(k+1)}-\frac{(\overline{\lambda}^{(k+1)}-\hat{\lambda}^{(k+1)})^2}{{\overline{\lambda}}^{(k+1)}-2\cdot\hat{\lambda}^{(k+1)}+\widetilde{\lambda}^{(k)}}\label{8c}\tag{8c}
\end{gathered}
\end{equation}
\subsubsection{Termination condition} Let $k\leftarrow k+1$ and repeat step 2) to 4) until converge conditions are satisfied, e.g.,
\begin{equation}
\begin{gathered}
\max\lbrace \mid\widetilde{\lambda}^{(k+1)}_{{\rm in},ji}(t)-\widetilde{\lambda}^{(k)}_{{\rm in},ji}(t)\mid\vert i\in\mathcal{I}, j\in\mathcal{I}_i\rbrace \leq \epsilon\label{9}\tag{9}
\end{gathered}
\end{equation}
where $\epsilon$ is a small positive value for convergence.
\section{Case study}
In this section, an case based on the IEEE 123-bus distribution network \cite{IEEE123} is studied. Two types of comparisons are simulated and analyzed in the case study to show the performance of the proposed method, one for comparison between the proposed real-time method and a pre-designed method, another for comparison between the proposed Aitken-DMPC solver and the standard DMPC (S-DMPC) solver, without any convergence acceleration.
\subsection{Settings}
In this case study, the length of one time step is $\Delta t$=10\,min and the length of the prediction window is 60\,min. Four subsystems are considered. The average travel speed of the repair crews is 30\,km/h when the scenarios of uncertainty and the influences from unpredictable events are not considered. The initial locations of the repair crews (location of the depots) can be determined by preparedness planning methods before the disasters, e.g.,\cite{arif2019stochastic}. The capacities of the DGs are 200\,kW, 250\,kW, 200\,kW, 250\,kW respectively. The value of $\varepsilon$ is 0.05 and the value of $\epsilon$ is 0.01.\\
\indent Besides, the flipping, swapping, and sliding probabilities are 0.3, 0.3, 0.3 respectively for the damage part of the chromosomes. The crossover and mutation probabilities are 0.3, 0.3 respectively for the crews part of the chromosomes, while 0.1 and 0.1 respectively for open-closed statuses part of chromosomes. The selection of the parent chromosomes for the next iteration is based on the tournament selection \cite{2000A}. Four parent chromosomes are selected and each parent chromosome generates 50 offsprings. Considering that the best chromosome over all the generations is always kept, there are 201 chromosomes in total for each generation. The number of generations is 50.\\ 
\indent In this section, 5 different cases with different damages and starting points of the repair crews are analyzed whose condensed information is given in Table I. The pre-designed method in \cite{7812566} is applied as comparison method. It is assumed that for the pre-designed method any newly emerging damage is assigned to be repaired when the pre-designed routes are all implemented, and that the repair crews do not alter their routes when traveling time changes caused by, e.g., blocked roads. As for the comparison between the solvers, the S-DMPC solver can be implemented in parallel in a similar way as the Aitken-DMPC but by replacing (8a)-(8c) by:\\
\begin{equation}
\begin{gathered}
\widetilde{\lambda}_{{\rm in},ji}(t)^{(k+1)}=\widetilde{\lambda}_{{\rm in},ji}(t)^{(k)}+\gamma_c(\widetilde{P}_{{\rm in},ji}(t)^{(k)}-\widetilde{P}_{{\rm out},ij}(t)^{(k)})\label{10}\tag{10}
\end{gathered}
\end{equation}
\begin{table}[!t]
\renewcommand{\arraystretch}{1.3}
\caption{Condensed information of other cases}
\label{table_example}
\centering
\begin{tabular}{cccccc}
\hline
Case & Number of damages & Number of crews & Number of depots\\ 
\hline
1 & 18 & 5 & 3 \\
2 & 14 & 5 & 3 \\
3 & 10 & 4 & 2 \\
4 & 8 & 3 & 2 \\
5 & 5 & 3 & 1 \\
\hline
Case & \tabincell{c}{Number of repair\\ duration changes} & \tabincell{c}{Number of traveling\\ time changes} & \tabincell{c}{Number of newly\\ emerged damages}\\
\hline
1 & 1 & 1 & 1 \\
2 & 3 & 2 & 0 \\
3 & 2& 3 &  1 \\
4 & 1 & 2 & 0 \\
5 & 0 & 2 & 1 \\
\hline
\end{tabular}
\end{table}
\indent It should be noticed that, to be concise, only the results of Case 1 are shown in detail, including explanations of the whole restoration process. The parameters and results of the other cases will be aggregated in Section V.C.
\subsection{The parameters and results of Case 1}
\indent In Case 1, at the beginning of the restoration process, there are 18 damages whose repair times are in Table II. Moreover, there are 5 repair crews, 4 DGs, and 5 switches as shown in Fig. 1. Besides, the following real-time conditions are considered. The maintenance duration for Damage 18 changes from 12\,min to 6\,min. The only path to Damage 10 is blocked at $t$=35\,min and is available again at $t$=45\,min. Damage 19 newly emerges at $t$=45\,min and 12\,min is needed to repair Damage 19. The routing results of Case 1 are indicated by the arrows in Fig. 5. The comparison results between the S-DMPC method and the proposed Aitken-DMPC method are shown in Fig. 6.\\ 
\begin{table}[!t]
\renewcommand{\arraystretch}{1.3}
\caption{Repair time for damages}
\label{table_example}
\centering
\begin{tabular}{cc|cc|cc}
\hline
Damage & \tabincell{c}{Repair\\time (min)}& Damage & \tabincell{c}{Repair\\time (min)} & Damage & \tabincell{c}{Repair\\time (min)}\\
\hline
1 & 15 & 7 & 9 & 13 & 9 \\
2 & 10 & 8 & 6 & 14 & 6 \\
3 & 18 & 9 & 6 & 15 & 15 \\
4 & 9 & 10 & 10 & 16 & 12 \\
5 & 9 &11 & 12 & 17 & 6 \\
6 & 12 &12 & 9 & 18 & 12 \\
\hline
\end{tabular}
\end{table}
\begin{figure}[!t]
\centering
\subfloat[The pre-designed routes of Case 1]{\includegraphics[scale=0.55]{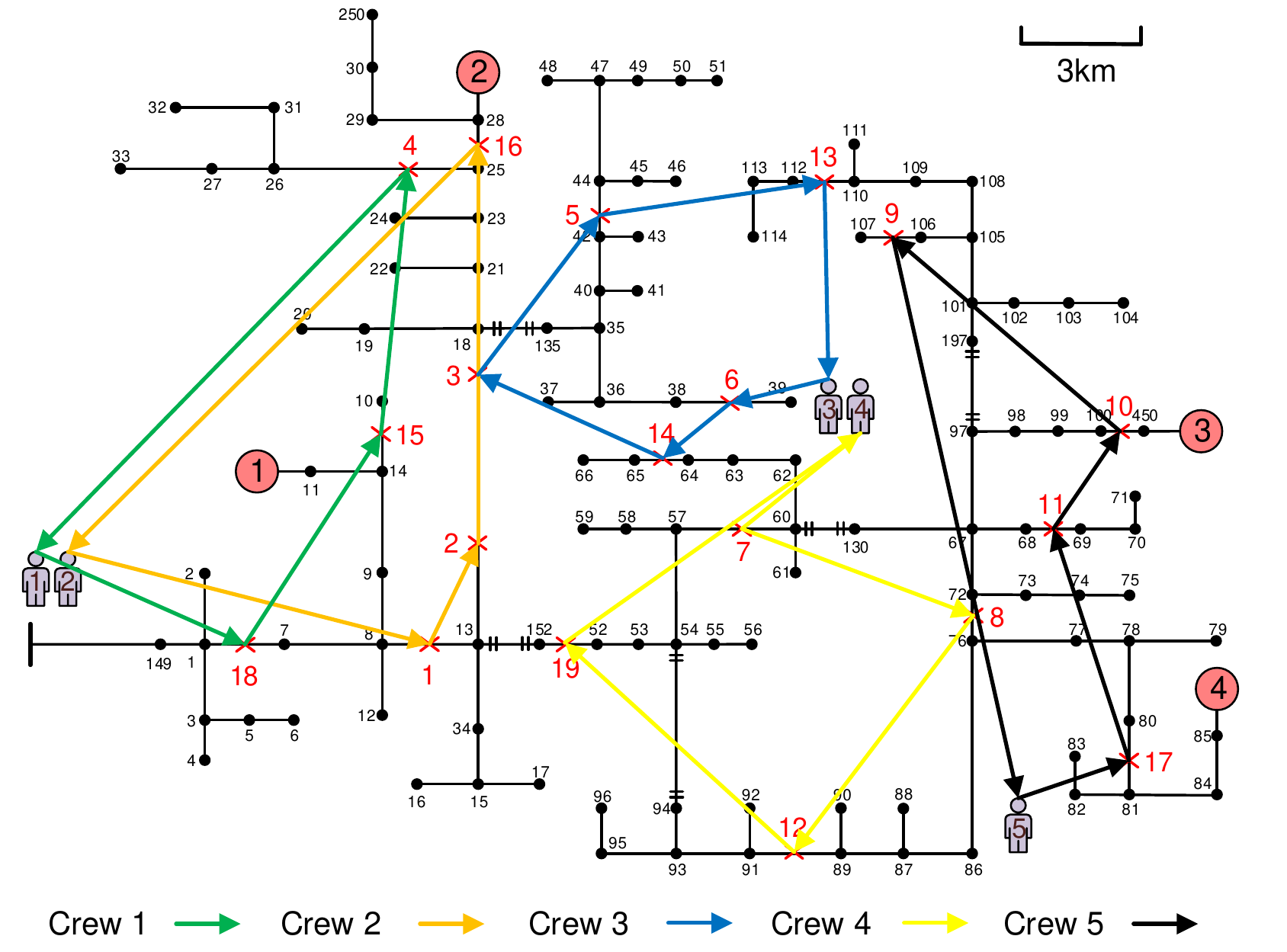}%
\label{fig_second_case}}
\hfil
\subfloat[The real-time routes of Case 1]{\includegraphics[scale=0.55]{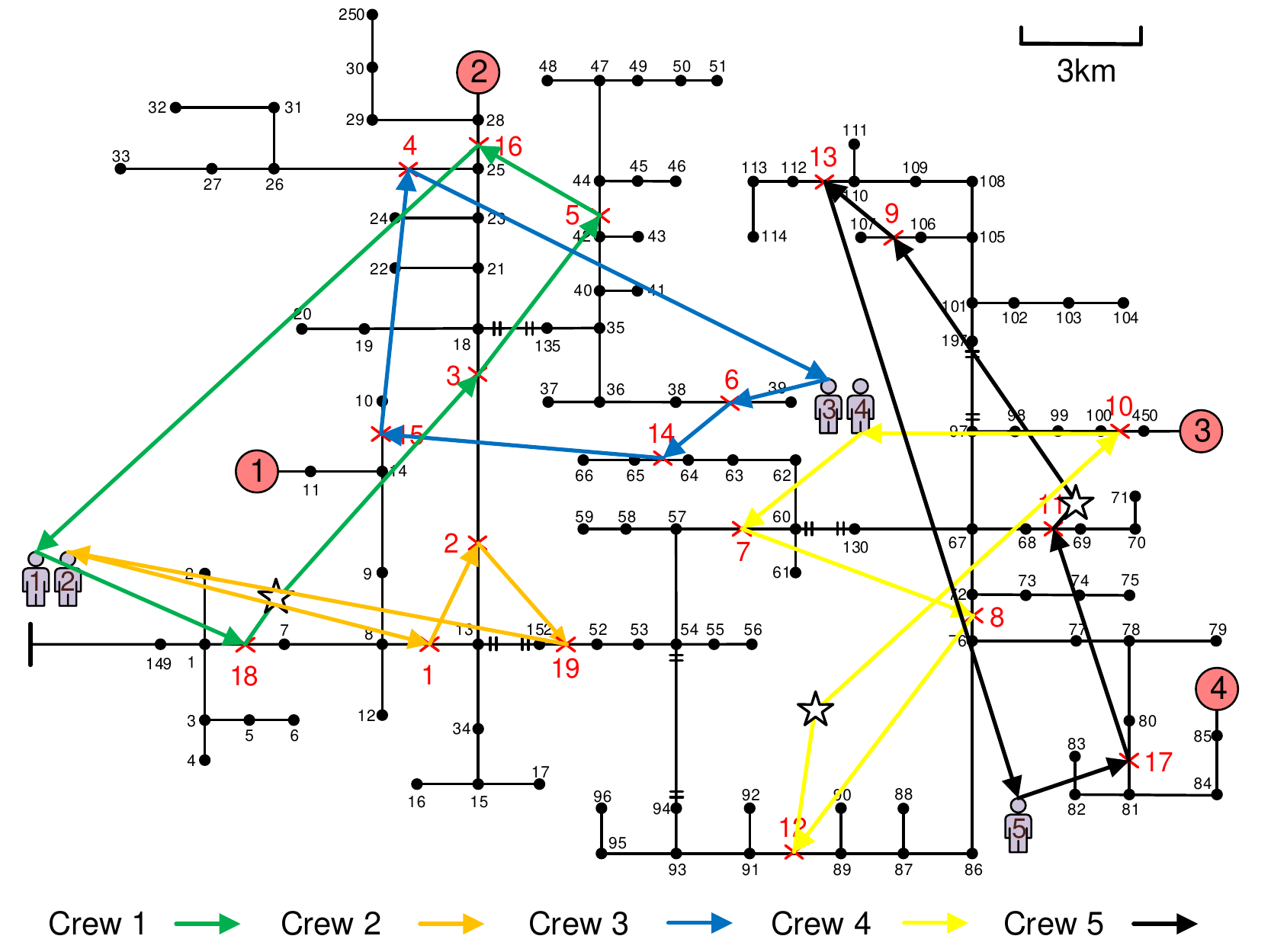}%
\label{fig_first_case}}
\caption{The comparison between the pre-designed plan and the Aitken-DMPC real-time plan}
\label{fig_sim}
\end{figure}
\indent In Fig. 5a, Crew 1 is allocated to repair Damage 18 and then at $t$=23.28\,min it has completed the repair and departs to Damage 15 in the pre-designed plan. However, in Fig. 5b, during the period from $t$=10\,min to $t$=20\,min, Crew 1 finds that only 6\,min is required to repair Damage 18 in the real-time situation. Then Crew 1 finishes repairing Damage 18 at $t$=17.28\,min and then departs to Damage 15. Crew 1 is at the location 5.22\,km away from Damage 15 (the hollowed star on the trace of Crew 1 in Fig. 5b) at $t$=20\,min. Then for the routing solution from the pre-designed strategy, Crew 4 is allocated to Damage 4 after repairing Damage 14. But for the routing solution in the real-time strategy, the power system operator redetermines the routes for $t$=20\,min to 30\,min according to the real-time locations of the crews. In the routing problem for $t$=20\,min to 30\,min, Crew 1 can reach Damage 4 earlier than Crew 4 because of finishing repairing Damage 18 earlier. Thus, in Fig. 5b, at $t$=20\,min, Crew 1 is allocated to Damage 3 and arrives at the location at $t$=36.25\,min. Crew 4 is allocated to Damage 15 and arrives at the location at $t$=43.6\,min.\\
\indent Then, between $t$=30 to 40\,min, the only path to Damage 10 is blocked. This real-time situation is included by setting the travel times to Damage 10 to a huge positive number, e.g., 1000\,min in this case study. In the pre-designed plan, before the blocking of this road, Crew 5 is allocated to Damage 10 after finishing repairing Damage 11 at $t$=37.64\,min and then waits until the road is available again. When performing the real-time strategy, at $t$=40\,min, Crew 5 is at the location of the hollowed star on the trace, and at $t$=40\,min the routes are rescheduled considering the blocked road to Damage 10. Thus, Crew 5 is allocated to Damage 9 and reaches it at $t$=57.2\,min.\\
\indent After that, between $t$=40\,min to 50\,min, Damage 19 newly emerges. In the pre-designed strategy, Damage 19 is addressed after all other damages have started to be repaired. Thus at $t$=60.81\,min, after finishing the repair of Damage 12, Crew 4 is allocated to Damage 19 and reaches it at $t$=73.97\,min. With the real-time strategy, Crew 2 is allocated to return to Damage 4 after finishing repairing Damage 2 at $t$=50.92\,min. When rescheduling the routes in real time at $t$=50\,min, Crew 2 is allocated to Damage 19 and reaches it at $t$=59.36\,min.\\
\indent Then, between $t$=60\,min to 70\,min, the only path to Damage 10 is available again. In the pre-designed strategy, Crew 5 now can access the location of Damage 10 and repair it. In the real-time strategy, before evacuating the path, Crew 4 is returning to its depot because all the damages are repaired or under repairing and is now at the location marked as the hollowed star on the trace of Crew 4 at $t$=70\,min (which is 7.23\,km away from the depot of Crew 4). At $t$=70\,min, the real-time routing strategy redetermines the routes of the current time step, and Crew 4 is allocated to Damage 10.\\
\indent For open-closed statuses of the switches determined by the real-time strategy, from $t$=50\,min to $t$=70\,min, the switch on 13-152 is open, and from $t$=80\,min to the end of the restoration process, the switch 60-130 is open. The other switches at other time steps are all closed. However, for the pre-designed strategy, from $t$=70\,min to the end of the restoration process, the switch 90-120 is open. While all other switches are closed.\\
\indent The load loss cost of the pre-designed plan during the restoration process is 1498\,\$, while the load loss cost of the real-time restoration plan computed with the proposed Aitken-DMPC approach is 1197\,\$. As can also be seen from Fig. 5, the proposed method can adapt to the changes of situations in real time to reduce load loss cost more than the pre-designed strategy. By determining the optimal routes step by step considering the changes in the real-time situation, the determined routes adjust to these changes smartly.\\ 
\indent Besides, the total computation time for the entire simulation has been compared and the results are shown in Fig. 6. In Fig. 6, the computation time reduction ratios for time step 1 to 8 are 18.4\%, 25.8\%, 5.8\%, 27.3\%, 25.7\%,	13.4\%, 9.5\%, 19.5\%. Thus, the proposed Aitken-DMPC approach can be used to accelerate the solution speed with respect to S-DMPC. Besides, the objective function values of these two approaches are within 3\% of each other. Furthermore, the GA algorithm has been run 10 times for each time step and for the results of the time step corresponding to $t$=$20$\,min, 9 in 10 runs yield the optimal solution obtained by mixed-integer linear programming solved which, however, is very time costly. While for the other time steps, 10 in 10 runs yield the optimal solution. So, the numbers of chromosomes and generations used in this case study are enough for GA to find the optimal solutions.   
\subsection{Results of other cases}
\indent Other cases have been tested associated with different numbers and locations of damages, different numbers of crews, different locations of depots, and different changes in real-time situations. The load loss costs of the pre-designed strategy and the real-time strategy for all the cases are listed in Table III. From Table III, it can be observed that the real-time strategy can reduce load loss more than the pre-designed strategy. Furthermore, the computation time and comparison between the proposed Aitken-DMPC method and S-DMPC are shown in Fig. 6. This figure shows that Aitken-DMPC is faster than S-DMPC in all the cases. The average reduction ratio of the computation time of all the cases of all the time steps is 18.55\%. Note that during the restoration process, the number of damages reduces due to the repair actions. As a consequence, the planning problem becomes less complex as the time progresses, and hence the computation time is also reduced.
\begin{figure*}[!t]
\centering
\includegraphics[scale=0.7]{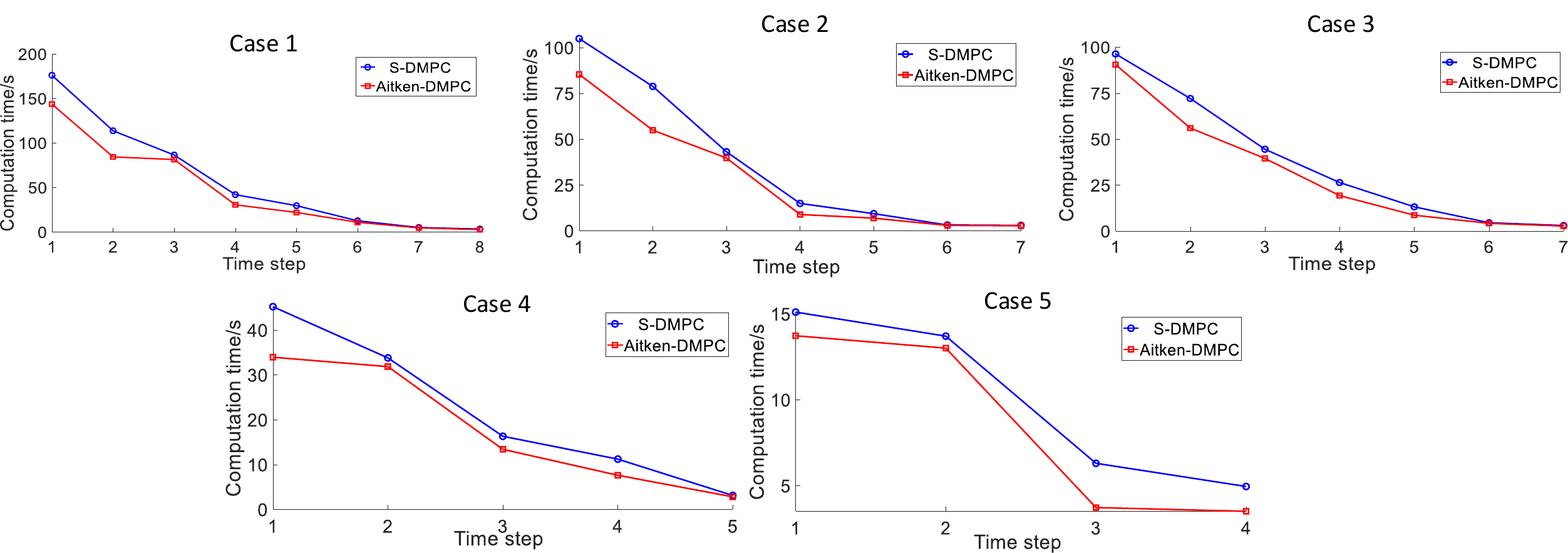}
\DeclareGraphicsExtensions.
\caption{Comparisons of solvers for five cases}
\label{fig_sim}
\end{figure*}
\begin{table}[!t]
\renewcommand{\arraystretch}{1.3}
\caption{Load loss cost comparison for Case 1 to 5}
\label{table_example}
\centering
\begin{tabular}{cccccc}
\hline
& Case 1 & Case 2 & Case 3 & Case 4 & Case 5\\
\hline
Pre-designed (\$) & 1498 & 1208 & 867 & 723 & 539 \\
Real-time (\$) & 1197 & 1002 & 699 & 649 & 476  \\
\hline
\end{tabular}
\end{table}
\section{Conclusions}
\indent This paper has proposed a real-time restoration planning strategy for distribution networks after disasters. Using the proposed strategy, the determined restoration plans, including routes of repair crews, reconfiguration, and power generation dispatch can adapt to the changes due to real-time situations. In this way, the load loss costs can be reduced during the restoration process. To determine the restoration plans in real time, a centralized-distributed bi-level optimization architecture has been proposed. By introducing a novel Aitken-DMPC solver, the proposed optimization architecture can solve the restoration planning problems in real time, and the solution speed is 18.55\% faster than the standard DMPC approach without convergence acceleration. Besides, regarding the final objective function values, the performance of Aitken-DMPC is comparable to standard DMPC without convergence acceleration.\\
\indent For future work, the preparation strategy before disasters will be studied, including determining the location of the depots for faster restoration, and hardening measures for vulnerable components.

\bibliographystyle{IEEEtran}
\bibliography{DMPC}\ 

\end{document}